# Predicted universality class of step bunching found on DC-heated Si(111) surfaces

A. Kozlov[1], A. Samardak[1], N. Chernousov[1], H. Popova[2], M.A. Załuska-Kotur[3], A. Pimpinelli[4], V.Tonchev[5, *]

[1]*Institute of High Technologies and Advanced Materials, Far Eastern Federal University, 10 Ajax bay Vladivostok, Russian Federation*

[2]*Institute of Physical Chemistry, Bulgarian Academy of Sciences, Acad. G. Bonchev str., block 11, 1113 Sofia, Bulgaria*

[3]*Institute of Physics, Polish Academy of Sciences, al. Lotników 32/46, 02-668 Warsaw, Poland*

[4]*Rice University, Houston, USA*

[5]*Faculty of Physics, Sofia University, 5 James Bourchier blvd., 1164 Sofia, Bulgaria*

* tonchev@phys.uni-sofia.bg

ABSTRACT. Concerted experimental and numerical studies of step bunching on vicinal crystal surfaces resulting from step-down electromigration of partially charged adatoms, confirmed the theoretical prediction of scaling dependence of the minimal bunch distance $l_{\min}$ on the bunch size $N$: $l_{\min} \sim N^{-\gamma}$, with $\gamma = 2/3$. The value of the so called size-scaling exponent $\gamma$ was observed in experiments on vicinal surfaces of semiconducting, metallic, and dielectric materials. Careful theoretical investigations and numerical calculations predict a second value of $\gamma$ - $\gamma = 1/2$. . However, this value is still not been reported from experiments. And we report here experimental observation of step bunching in the universality class relative to $\gamma = 1/2$. This is achieved by monitoring step flow during sublimation of Si(111)-vicinals heated by a direct step-down current at ~1200°C. In the experiment we also measure other characteristic for the bunching quantities, such as the mean total number of steps in the bunch $N$ and the mean bunch width $W$. We then compare our findings with published experimental and numerical data to arrive at a theoretically consistent framework in terms of universality classes. The ultimate benefit of our study is not only to advance fundamental knowledge but also to provide further guidance for bottom-up synthesis of vicinal nanotemplates.

KEYWORDS: nanotemplates, step bunching, vicinal crystal surfaces, electromigration, scaling and universality, defect engineering,

Modern theory of crystal growth starts with the identification by Kossel [1] and, independently, by Stranski [2] of the so-called *half-crystal* or *kink* position, where an atom—or the building unit composing the crystal—has an energy equal in absolute value to half the sublimation energy per particle of the whole crystal; the latter can therefore grow along the lowest entropy production pathway when building units attach to kinks; see also [3]. Macroscopically, this implies that the crystal can grow towards a self-similar polygonized shape. In this picture the defects turn out to be essential for the growth providing an array of the necessary growing sites-– the surface steps.

These are always present when a crystal is cut along a high symmetry direction, because the cut will necessarily always form a (small) angle ("miscut angle") with respect to the high-symmetry, or low-index, surface orientation. The result is a vicinal (*neighboring* the high symmetry orientations) surface, on which steps appear where the cut plane intersects the crystal lattice. Vicinal crystal surfaces are convenient templates [4] to grow various homo-and heteroepitaxial layers and nano-structures [5–9]. However, step edges effectively move during growth, and because of asymmetries and anisotropies in the various processes by which building units diffuse and attach to the steps, the initial approximately uniform distribution of steps is often unstable and is lost as the growth proceeds. Even when asymmetries are not intrinsic to the incorporation process of the building units, recent theoretical developments have shown that step motion itself is a source of instability [10–12].

Another extrinsic phenomenon that affects the morphology of crystal surfaces is the electromigration, which is a source of failure in many electronic circuits [13]. Its controlled study has made use of crystalline sample with vicinal (stepped) surfaces; when a current flows through the bulk, surface steps can be observed to bunch or to be made equidistant, depending on the direction of the current with respect to the direction of the step staircase. On Si crystals, where most of the investigations were and are being performed, the instability is believed to be a consequence of the action of the applied electric current on surface Si atoms (adatoms).

Within the so called B2-type of step bunching [14] the coarsening of the surface morphology can be quantitatively characterized by measuring the average number of steps in a bunch $N$, the average bunch width $W$, the minimum distance $l_{min}$ between steps in a bunch. These quantities are observed to be increasing in time as power laws: $N \sim t^\beta$, $W \sim t^{1/z}$, $l_{min} \sim t^{-\delta}$, which define the dynamic exponents $\beta, z$ and $\delta$. Furthermore, the morphology can be seen to obey scaling relations of the form $W \sim N^{1/\alpha}$, $l_{min} \sim N^{-\gamma}$, with the static exponents being $\alpha = \beta z$ and $\gamma = 1 - 1/\alpha$.

Because of the ubiquity and variety of morphological instabilities, of the importance of such examples of self-organization, and partly inspired by investigations of out of equilibrium systems epitomized by the celebrated Kardar-Parisi-Zhang equation [15], a unified description of the bunching of straight steps was proposed by Pimpinelli et al. [16] (PTVV) and then generalized by Krug et al. (KTSP) [17]. These authors argued that the asymptotic morphology of an unstable vicinal surface should generically obey the following partial differential equation:

$$\frac{\partial h}{\partial t} + \frac{\partial}{\partial x}\left(K_1 m^\rho + \frac{K_2}{m^k}\frac{\partial^2}{\partial x^2} m^n\right) = const \qquad (1)$$

where $h$ is the surface height, $x$ is the spatial coordinate, $K_1$ and $K_2$ are material parameters that summarize the details of the system-specific destabilizing kinetics ($K_1$), and of the step-step energetic interactions ($K_2$), respectively; $m \equiv \partial h/\partial x$ is the local surface slope. The exponents $\rho$ and $k$ describe the destabilizing role of the kinetics, assuming that the latter only depends on the local surface slope. Finally, $n$ determines the interaction between steps at distance $d$, $U \sim 1/d^n$ where linear elasticity predicts $n = 2$. Eq. (1) allows one to compute the morphological exponents $\alpha, \beta$, etc. as a function of the kinetic parameters $\rho$ and $k$. Thus for static exponents we have $\alpha = 1 + 2/[n - (k + \rho)]$ and $\gamma = 2/[2 + n - (k + \rho)]$. Note that $\rho K_1 > 0$.

| ρ | -1 | | | | | 0 | | | | | 1 | | | | |
|---|---|---|---|---|---|---|---|---|---|---|---|---|---|---|---|
| k | δ | γ | β | α | z | δ | γ | β | α | z | δ | γ | β | α | z |
| 0 | 1/5 | 2/5 | **1/2** | 5/3 | 10/3 | **1/3** | **1/2** | **2/3** | **2** | **3** | 1 | **2/3** | 3/2 | 3 | 2 |
| 1 | 1/4 | 1/2 | **1/2** | 2 | 4 | 1/2 | **2/3** | 3/4 | 3 | 4 | | | | | |
| 2 | 1/3 | **2/3** | **1/2** | 3 | 6 | | | | | | | | | | |

Table 1. The values of the scaling exponents as derived from the dimensional analysis of Eq. (1) with fixed value of *n*=2. Exponents are defined in the body of the text. The column in green contains constant values of $\beta$ =1/2, the "diagonal" in red – constant values of γ. These values are simultaneously found only when *k*=2; see also Krzyżewski et al. *[18]* . The exponents in the blue-shaded cells were found in a numerical study *[19]* and experimentally in the present work.

Table 1 contains the value of the exponents as predicted by Eq. (1) as a function of the kinetics, for elastic interactions with $n = 2$. The main merit of Eq. (1) is to reveal the existence of step bunching "universality classes", similar to equilibrium critical phenomena, both for the experiments and the atomistic models that are designed to describe unstable step motion in specific material systems. Eq. (1) then makes explicit the underlying, generic structure of the destabilizing mechanisms.

Inspection of Eq. (1) suggests that universality classes can be labeled by the value of $\rho + k$ . In particular, we will discuss here two such classes: class (0) corresponding to $(\rho + k) = 0$, and class (1) corresponding to $(\rho + k) = 1$. For the physically relevant case $n = 2$, class (0) is characterized by static exponents $\alpha = 2, \gamma = 1/2$; class (1), by $\alpha = 3, \gamma = 2/3$. Each class contains further subclasses depending on the value of $\rho$, which sets the value of the dynamical exponents $\beta$ and $z$; for instance the product $z(\beta - 1/2) = (\rho + 1)/(n - \rho - k)$, so that $\rho = -1$ implies $\beta \equiv 1/2$ for all $n$, as well as $z = 2\alpha$. Therefore, if $\rho = -1$ and $n = 2$: in class (0) $z = 4$ , while in class (1) $z = 6$. As a corollary, one must conclude that measuring only static $(\alpha, \gamma)$ or dynamic $(\beta, z)$ exponents is not sufficient to assign a system or a model to the appropriate universality class: at least one of each type of exponents is needed; as a matter of fact, because of the unavoidable experimental or numerical uncertainties affecting the exponents measured in experiments or computed in models, as many exponents as possible should be directly measured or computed, respectively.

Step bunching induced by electromigration on vicinals of Si (111) has become the paradigmatic playground in which to investigate this type of morphological instability. Silicon is an extremely well known material, which allows investigators to be very confident that experiments are probing intrinsic properties in a completely reproducible way. Moreover, the observed behavior, as a function of surface orientation, temperature, current amplitude and direction with respect to the steps ("step-up" vs. "step-down" current) is complex enough to present yet unresolved challenges to its full understanding (see e.g. Fig. 3 of Ref. [20]). So far, several experimental groups have studied this instability [20–23]; they have concentrated their efforts on static exponents in the temperature regimes where step bunching occurs with a step-down current; they have consistently reported $\gamma = 2/3$, implying that this system belongs to the the universality class (1) ($\rho + k = 1$). When possible, they have measured $\beta$, and consistently found $\beta = 1/2$. Together, these results are consistent with the predictions of Eq. (1) with $\rho = -1, k = 2$. Not surprisingly, most published models also fall in this class [24,25]. There is however evidence from model calculations [17,19,26], as well as from experiments [22]—which we will discuss below—that, depending on the experimental parameters, electromigration induced step bunching on vicinals of Si(111) with step-down current might belong to *either* of the (0) or (1) universality classes. Later, $\gamma = 2/3$ was found also on metallic – W(110) [27] and dielectric – Al2O3 [28] surfaces.

In this work, we report the first clear and complete experimental observation of step-down, electromigration-induced step bunching belonging to the ($\rho + k = 0$) universality class, with exponents coinciding with the predictions of Eq. (1) for $\rho = k = 0$.

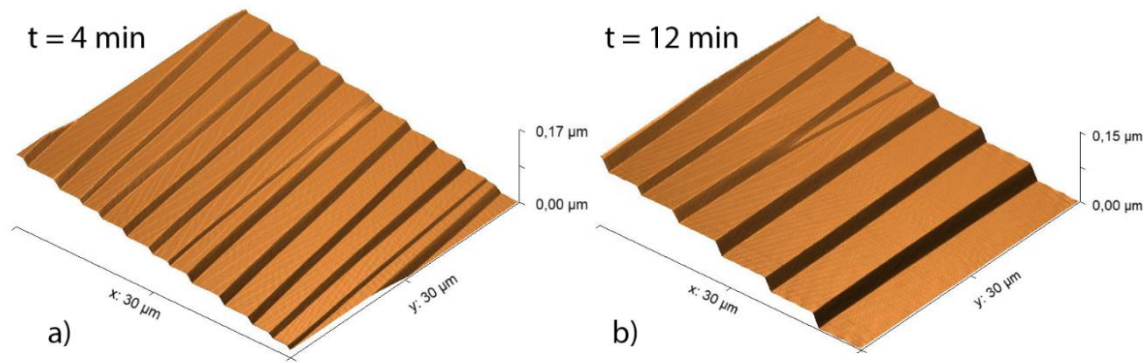

Fig.1. AFM images of the sublimating Si(111) vicinal surface, misoriented in the [11-2] direction by 0.3º, heated by step-down direct current for 4 min (a) and 12 min (b).

We used Sb-doped vicinal single-crystal Si(111) substrates with resistivity of about 10 *Ohm cm* misoriented along [1 1 -2] by 0.3° and 2°, respectively. After chemical cleaning and indirect baking degassing, samples were annealed by direct current at 1200ºC in an ultrahigh vacuum chamber with base pressure of approximately 4×10$^{-11}$ Torr. The temperature was measured with an optical pyrometer. Step-down DC heating was performed within 4-60 min, after which the surface morphology was measured with scanning tunneling microscopy at ultrahigh vacuum. The STM study was carried out for profile analysis of step-bunched surfaces and to check the step-down direction of the vicinal surface. Scanning was carried out in the constant tunneling current mode, the value of which was 1 nA, and the bias was 2 V. Atomic force microscopy (AFM) was used at the normal conditions in semi-contact mode to estimate the step-bunch distribution on the crystal surface over a larger area than that accessible with STM. Fig. 1 shows two AFM images of the sublimating Si(111) vicinal surface. They reveal a surface morphology consisting in regular, straight step bunches formed after 4 (Fig. 1a) and 12 (Fig. 1b) minutes of heating by step-down direct current, respectively. Image processing and profile analysis were performed to determine average bunch parameters (bunch size, bunch width, average and minimum step-step distance within a bunch, terrace width, etc.). This approach was described in [29] where it is called «Monitoring Scheme I».

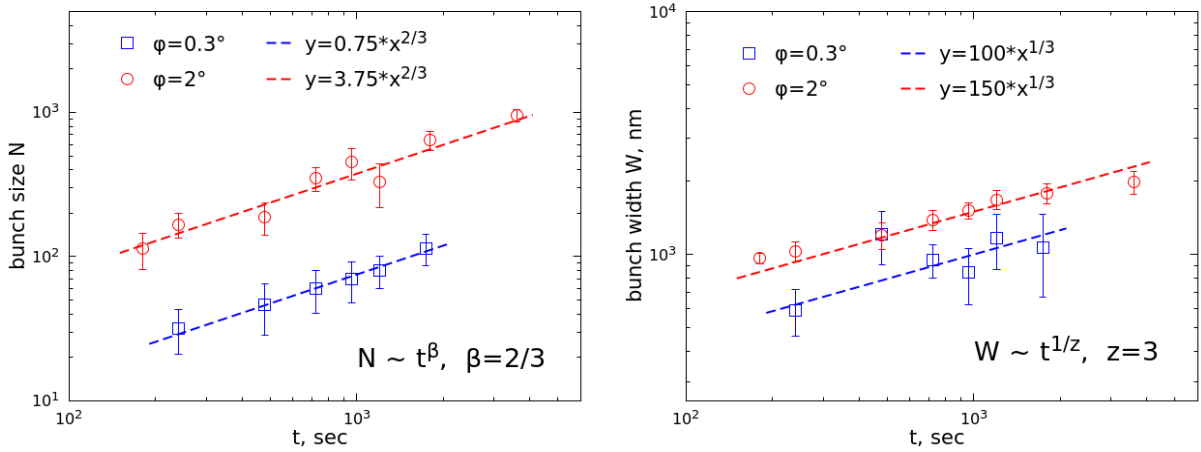

Fig.2. Bunch size $N$ (defined as the number of steps in the bunch; left panel) and bunch width $W$ (right panel) as a function of the annealing time $t$, for two vicinal surfaces of Si(111) misoriented by 0.3º (blue squares) and 2º (red circles). The obtained time dependences, together with the corresponding scaling exponents, are shown in the insets.

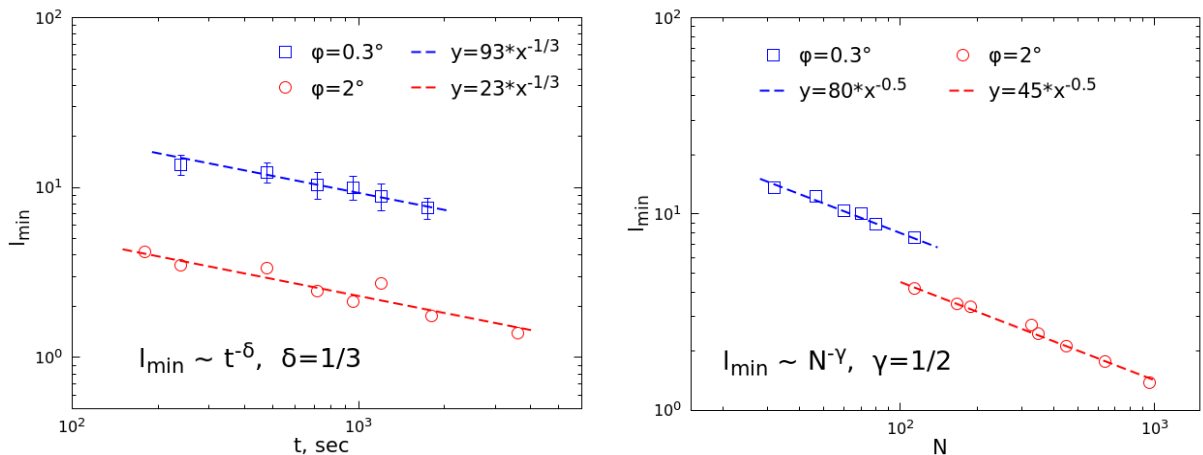

Fig.3 Scaling of the minimal bunch distance $l_{min}$ with time $t$ (left panel) and with the bunch size $N$ (right panel) for vicinal surfaces of 0.3º (blue squares) and 2º (red circles).

Figures 2 and 3 show the results of this analysis. It can be seen that the measured scaling exponents for both the 0.3° and 2° miscuts coincide with the exponents given in Table 1 for $k=0$ and $\rho=0$. In addition, Fig. 4 shows that the scaling exponent of $l_{min}$ vs $N$ is $\gamma=1/2$, confirming that this system belongs to the universality class (0). The latter result is compared in Fig 4 with numerical calculations from Ref. [19]. We have also reconsidered results previously reported in Ref. [22]. In that work, the authors studied electromigration-induced step bunching on Si[111] with step-down current, and either advancing (net growth) or retreating (net sublimation) steps. They report measurements of $l_{min}$ and of $N$ as functions of the average terrace width $l_0$ (proportional to the reciprocal of the tangent of the miscut angle.) By eliminating $l_0$ from their data measured at T=940°C, we have plotted the latter in Fig. 4 as $l_{min}$ vs $N$. Although the corresponding bunch sizes are rather small, we found, for both growth and sublimation conditions, the same scaling exponent $\gamma=1/2$ as in our experiment. One should note that proceeding in the same way with the other measurements of $l_{min}$ and $N$ reported in [22] for higher temperatures (not shown here), we found instead $\gamma=2/3$.

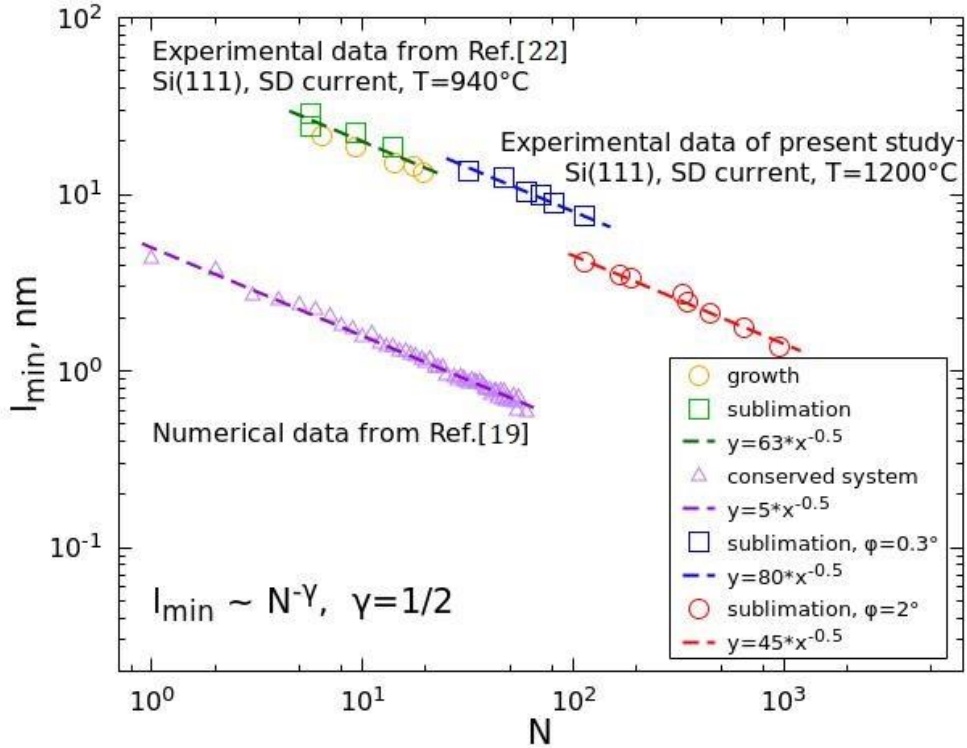

Fig.4. Comparison between experimental data for the size-scaling dependence of $l_{min}$ presented in this study and previously published numerical data in [19], but re-plotted with taking the value of vicinal angle 2°, and experimental data in [22]. SD is for step-down.

The experimental results reported in this work, as well as those previously published, lead us to conjecture that the step bunching resulting from the electromigration-induced instability is driven by a surface current of the form

$$J(m) = \frac{K_1}{1+d_1 m} + \frac{K_2}{1+(d_2 m)^2} \frac{\partial^2}{\partial x^2}(m^2) \qquad (2)$$

where $d_1$ and $d_2$ are characteristic lengths associated with step kinetic processes (atom attachment/detachment, step motion, etc.) Eq.(2) then shows how changing the experimental conditions affecting $d_1$ and $d_2$ can modify the scaling of the surface profile from $\rho = k = 0$ when $d_1 m \ll 1$ and $d_2 m \ll 1$, to $\rho = -1, k = 2$ when $d_1 m \gg 1$ and $d_2 m \gg 1$. The crossover would happen when $d_1 \sim d_2 \sim l_{min} \sim m_{max}$. Note that $d_1 m \to 0$ implies $K_1/(1 + d_1 m) \sim K_1(1 - d_1 m) \sim K_1 ln(d_1 m)$, which leads to the conclusion that $\rho = 0$ in fact means a weak, logarithmic dependence on the slope.

One should notice, though, that in an attempt to reproduce experimental results reported for step bunching during growth [30] Jabbour and coworkers have found [12] $\beta = 1/2, \gamma = 2/3$, corresponding to $\rho = -1, k = 2$. Their model maps onto a continuum equation similar to Eq. (1) above, albeit with leading order terms $\rho = -1 \ and \ k = 1$. It is not clear whether their model does not satisfy self similarity, or if divergences in sub-leading term contributes to modify $k = 1$ into an effective parameter k = 2.

In reporting here for the first time on the experimental observation of coarsening exponents $\beta = 2/3, \gamma = 1/2$ during step bunching, we showed that Si electromigration belongs to a type (0) universality class for appropriate values of the control parameters, and to a type (1) class for different values. This observation complicates further an already complicated picture. The morphological phase diagram of Si (111) established so far (see fig.3 in [20]) exhibits four different step bunching regions, depending on temperature T and current direction (SD or step-down; SU or step-up): I (SD, 860<T<960°C); II (SU, 1060<T<1200°C); III (SD, 1200 <T<1300 °C); IV (SU, T>1320 °C). Between regions I and II, Si (111) vicinals are morphologically stable (equidistant steps) for both SD and SU conditions. Before the work reported here, only exponents corresponding to type (1) universality, $\beta = 1/2, \gamma = 2/3$, had been measured for SD conditions (regions I and III); in this work we have shown that at least close to the lower limit of region III, step bunching exhibits type (0) universality, $\beta = 2/3, \gamma = 1/2$. We have also discussed (not completely conclusive) evidence that type (0) universality may also occur in region I. It is therefore necessary that experimentalists meet the challenge to reassess the scaling behavior of Si(111) vicinals by measuring as many exponents as possible, not just $\beta \ and \ \gamma$. On the other hand, theorists must address the even greater challenge of deriving more microscopic (discrete) models based on a more realistic description of electromigration, accounting e.g. for step motion that *advects* the adatoms [12], and possibly for the effect of the electric field on step-step interactions. The scaling of the stabilizing terms that comes from step-step interactions in Eq. (1) is also an outstanding issue. As shown in [17], the exponent $k$ is expected to be either 1 or 0 only. In order to have type (1) universality with $\beta = 1/2, \gamma = 2/3$, $k$ must equal 2. We can surmise that the origin of this scaling lies in the out-of-equilibrium adatom concentration near the step edge. This interpretation is suggested by the scaling behavior of the $c^+ - c^-$ one-dimensional model [29], introduced as an example of a model exhibiting step bunching during growth and possibly exhibiting meandering in higher dimensions. The scaling behavior of its bunching instability can be found from Eq. (1) provided one sets $\rho = -1, k = 1 - n$. In the $c^+ - c^-$ model, the value of $k$ is affected by non-equilibrium effects, similarly to what is observed in electromigration. Supporting arguments to this suggestion can be found in [12], where a rigorous approach to step flow growth was developed that shares some elements (called by those authors the "chemical effect"; see [12] for details) with the $c^+ - c^-$ model. The authors computed scaling exponents for coarsening of the step bunches, and found values in agreement with $= -1, k = 2$, that belong to a type (1) universality class. More work needs to be done to clarify the points above, and to approach a better understanding of the possible universality classes for coarsening during step bunching.

VT acknowledges Mercator Fellowship from the German Research Foundation (DFG) at the Federal Institute for Materials Research and Testing (BAM) and thanks for the hospitality of the Glass Division there. A.S.S. and A.G.K. acknowledge the support of the Russian Ministry of Science and Higher Education under the state task No. 0657-2020-0013.